Two-dimensional Ising model in a finite magnetic field for the square lattice of infinite sites-Partition function and Magnetization

M V Sangaranarayanan

Department of Chemistry

Indian Institute of Technology Madras

Chennai-600036   India

sangara@iitm.ac.in

Abstract

The partition function and magnetization equations are derived for a two-dimensional nearest neighbour Ising models for a square lattice. The formulation involves exploiting Onsager's exact solution as an input in conjunction with the universality hypothesis.

1. Introduction

Despite innumerable attempts at solving the two-dimensional nearest neighbour Ising models exactly in a finite magnetic field, the outcome has largely remained elusive. In hindsight, this is surprising since the analysis for zero magnetic field case wherein interesting critical phenomena occur has been elegantly analysed by Onsager[1],Lee and Yang[2], Kramers and Wannier [3] Kac and Ward[4],Wilson [5] and Montroll [6].The computational efforts not withstanding, convenient analytical expressions for partition functions and thermodynamic functions derived therefrom are lacking if external magnetic field is introduced into the Ising Hamiltonian. Among several methods amenable for deducing the partition function and magnetization for the two-dimensional nearest neighbour Ising models, mention may be made of the following:(i)Construction of Toeplitz matrices for finite lattices etc[7];(ii) graph theoretical approach of counting the number of black-white edges [8]followed by asymptotic extrapolation;(iii)Incorporation of magnetic field into the partition function equation using the random walk perspective[9] etc.

In this Communication, the partition function as well as magnetization for a square lattice of infinite sites, pertaining to two-dimensional nearest neighbour Ising models is reported employing the scaling and universality hypothesis in conjunction with a graph theoretical procedure.

2. Methodology

The scaling and universality hypothesis has played a central role in the development of the modern theories of critical phenomena[10]. While the estimation of critical temperatures has attracted a great deal of attention and served as a touchstone to investigate the efficacy of approximations such as Bragg-Williams approximation[11],Bethe *ansatz* [12], high and low temperature series expansions[13] etc, the influence of various critical exponents on the magnitude of the appropriate partition functions is not obvious. In order to remove this lacuna, it is imperative to deduce the partition function itself in terms of critical temperatures and critical exponents. For this purpose, the exact solution of Onsager is employed as an *input* to derive the canonical partition function in a non-zero magnetic field.

Consider the two-dimensional nearest neighbor Ising model on a square lattice with the Hamiltonian given by

$$H_T = -J \sum_{<ij>} (\sigma_{i,j}\sigma_{i,j+1} + \sigma_{i,j}\sigma_{i+1,j}) - H \sum \sigma_{i,j}$$

where J denotes the nearest neighbor interaction energy, H being the external magnetic field, i and j denote the row and column index respectively.

(A)Partition function

As has been earlier shown, the counting of black-white edges leads to the partition function for a square lattice of finite sites[8]. These energetic contributions can then be employed judiciously to derive the approximate critical temperatures. Employing the properties of Gauss Hypergeometric series [14] and an identity given by Ramanujan [15], the partition function Q for a square lattice of N sites follows as

$$\frac{1}{N}\ln Q = \left(\frac{T_c}{T}\right)\frac{1}{8}\left[-M^8 - \log(1-M^8)\right]$$
$$+ \left(1-\frac{T_c}{T}\right)^{7/4}\frac{1}{8}\left[\log\frac{M^2+1}{M^2-1} - 2\tan^{-1}\frac{1}{M^2}\right] + constants$$

(1)

where M denotes the magnetization, T is the absolute temperature, $T_c$ being the critical temperature. The constant terms are independent of the external magnetic field 'H' and ensure that the Onsager's zero field partition function equation is recovered when H =0. When the external magnetic field is zero, M becomes $M_0$, the spontaneous magnetization and is given by the familiar Lee and Yang eqn[2]. It may be pointed out that the occurrence of the critical temperature ($T_c$) and the magnetization critical exponent (1/8) within the partition function of two-dimensional nearest neighbour Ising models is entirely new. The above eqn also enables the estimation of the critical exponent corresponding to the specific heat, magnetization, susceptibility etc.

(B) Magnetization

The magnetization for the lattice of infinite sites can be deduced from the above partition function effortlessly. However, the resulting eqn is transcendental and requires inversion such as Lagrange's inversion [16] and is not easy to decode. *In view of this*, an approximate equation for the magnetization is derived and is as follows:

$$M \simeq 2h_t^{1/15} - M_0 + \frac{M_0^{14}\left[h_t^{-13/15} - M_0^{-13}\right]}{13(M_0^8 - 1)^{11/14}} - \frac{1}{9}\left[h_t^{9/15} - M_0^9\right] + \text{higher order terms}$$

(2)

where

$$h_t \simeq \frac{15H}{kT_c} + M_0^{15} \quad (3)$$

As a verification of the validity of eqn (2), it can be seen that (i) when the magnetic field is zero, M →$M_0$ consistent with the Lee and Yang eqn [2];(ii) when T = $T_c$ , the critical exponent for non-zero magnetization M ~ $H^{1/15}$ [6];(iii) the zero field magnetic susceptibility can be shown to diverge at the critical temperature and (iv) with inclusion of the higher terms in eqn (2) or employing the exact algebraic eqn, the magnetization turns out to be an odd function of the magnetic field. In view of the equivalence between the Ising model and lattice gas description, eqns (1) and (2) have profound implications in surface science and condensed matter physics.

3. Summary

The magnetization and partition functions for two-dimensional nearest neighbour Ising models for square lattices in a finite magnetic field is reported using the Onsager's exact solution as an input in conjunction with enumeration of the black-white edges in a square lattice.